\documentclass[pdflatex,sn-mathphys]{sn-jnl}

\jyear{2021}%

\theoremstyle{thmstyleone}%
\theoremstyle{thmstyletwo}%

\theoremstyle{thmstylethree}%

\usepackage{multicol}
\usepackage{lipsum}
\begin{document}

\title[Verifi-Chain: A Credentials Verifier using Blockchain and IPFS]{Verifi-Chain: A Credentials Verifier using Blockchain and IPFS
}


\author[1]{\fnm{Tasfia} \sur{Rahman}}\email{tasfia.rahman.cse@ulab.edu.bd }
\equalcont{These authors contributed equally to this work.}

\author[1]{\fnm{Sumaiya Islam} \sur{Mouno}}\email{sumaiya.islam.cse@ulab.edu.bd }
\equalcont{These authors contributed equally to this work.}

\author[1]{\fnm{Arunangshu Mojumder} \sur{Raatul}}\email{arunangshu.mojumder.cse@ulab.edu.bd}
\equalcont{These authors contributed equally to this work.}

\author[2]{\fnm{Abul Kalam} \sur{Al Azad}}\email{abul.azad03@northsouth.edu}
\equalcont{These authors contributed equally to this work.}

\author[1]{\fnm{Nafees} \sur{Mansoor}}\email{nafees.mansoor@ulab.edu.bd }
\equalcont{These authors contributed equally to this work.}

\affil*[1]{\orgdiv{Computer Science \& Engineering}, \orgname{University Of Liberal Arts Bangladesh (ULAB)}, \orgaddress{\city{Dhaka}, \country{Bangladesh}}}

\affil*[2]{\orgdiv{}, \orgname{North South University}, \orgaddress{\city{Dhaka}, \country{Bangladesh}}}


\abstract{Submitting fake certificates is a common problem in Southeast Asia, which prevents qualified candidates from getting the jobs they deserve. When applying for a job, students must provide academic credentials as proof of their qualifications, acquired both inside and outside the classroom. Verifying academic documents before hiring is crucial to prevent fraud. Employing blockchain technology has the potential to address this issue. Blockchain provides an electronic certificate that is tamper-proof and non-repudiable, making it difficult for students to manipulate their academic credentials. This paper presents a prototype for an academic credential verification model that leverages the security features of blockchain and IPFS (Interplanetary File System). Certificates are temporarily stored in a database before being transferred to IPFS, where a unique hash code is generated using a hashing algorithm. This hash code serves as the certificate's unique identity and is stored in the blockchain nodes. Companies can verify an applicant's credentials by searching for the applicant and accessing their already verified certificates. Utilizing IPFS as a middleman storage platform lowers the expenses of directly storing massive data on the blockchain. To sum it up, the proposed solution would make the process of certificate verification more efficient, secure, and cost-effective. It would save time and resources that would otherwise be used to manually verify certificates.}




\maketitle

\section{Introduction}\label{sec1}

The basic structure of the mainstream education system includes primary, secondary, and tertiary \cite{paraide2023before}. Hence, after the completion of primary and secondary education, students get enrolled in universities and can pursue further studies based on their preferences. Furthermore, students participate in a variety of extracurricular activities throughout their academic years. This means that they receive a plethora of certificates throughout each stage of their education journey. The problem regarding this situation is that these certificates are at an increased risk of being lost or damaged, and there is no regulated system in place to store all these certificates digitally or verify their authenticity.

Many countries have huge populations, and with millions of graduates applying for jobs each year, individually verifying the credentials can be extremely time-consuming and taxing. It is exceedingly challenging to manage and authenticate such a vast amount of records, leading to an unfavorable situation where falsified or replicated certificates can be created through tampering. This aspect has given rise to an increasing number of fraudulent organizations that have been engaging in the unethical practice of forging academic certificates. Unfortunately, as technology advances, it has become more difficult to distinguish between genuine and forged certificates \cite{gopal2018survey}.

To address this issue, this proposed system utilizes Blockchain, a new emerging sophisticated technology. So, what are the benefits of using Blockchain? The immutability and tamper-proof nature of blockchain make it a very robust system to use \cite{monrat2019survey}. Even if the state of the data is compromised, it can detect the change in less than a second. In Blockchain, data or nodes are validated only when multiple parties approve them \cite{zhang2019security}. As a result, the system would always be reliable and authenticated. It is not only secure but extremely transparent about the transactions occurring in the system and there is also a traceability aspect to it.

Blockchain technology has rapidly gained popularity in recent years as a novel and promising approach to securely store, share, and manage data. Originally developed as a distributed ledger technology to support cryptocurrencies such as Bitcoin, blockchain has now evolved to become a versatile and robust platform for a wide range of applications beyond finance, including supply chain management, healthcare, real estate, voting systems, and more \cite{chatterjee2017overview}. At its core, blockchain technology offers a decentralized and immutable database that is resistant to tampering and fraud. It achieves this by employing a consensus mechanism that ensures the integrity of the ledger and eliminates the need for intermediaries or central authorities to manage the data. This makes blockchain technology an ideal solution for use cases that require high levels of trust, transparency, and security.
By using IPFS, files can be stored and accessed in a secure, decentralized, and censorship-resistant way \cite{benet2014ipfs}. The integration of IPFS with blockchain technology provides an additional layer of security and immutability to the file storage system \cite{kumar2019implementation}. Since everything is stored digitally and all the certificates are verified before being stored in the IPFS, students do not have to worry about losing or damaging their certificates. Furthermore, it streamlines the process for companies to view these verified certificates and hire eligible applicants accordingly. As a result, this proposed system closes the gaps in the current system and provides us with an effective and tangible solution.

The proposed solution aims to design and implement a decentralized certificate verification system using Blockchain and IPFS technology. The system enables secure and tamper-proof storage of digital certificates, allowing for easy and efficient verification of the authenticity and integrity of the certificates by authorized parties.
By utilizing Blockchain technology, data stored on the network is guaranteed to be unchangeable and transparent. Additionally, IPFS allows for decentralized and distributed storage of certificates. The system also incorporates a user-friendly interface for convenient access and utilization.

The objective of the paper is to chart the present status of understanding regarding Blockchain and IPFS. After the constructive Abstract and Introduction, there is a Literature review in section 2 along with this Proposed system in section 3. In section 3 there are some subsections that briefly describe the system module, architecture, and so on. Ultimately, the intention of this paper is to illuminate the current condition of the verification process and suggest a strong system for future implementation.

\section{Related Work}\label{sec2}

The primary objective of this literature review is to delve into the current state of knowledge regarding academic certificate verification using blockchain technology, taking into account a multitude of factors such as various approaches, strategies, methodologies, and techniques employed in previous research studies, with the ultimate goal of gaining a comprehensive understanding of the field and identifying potential areas for further investigation and improvement.

A thorough examination of the application of blockchain technology in smart contracts, including the basic principles of blockchain, such as decentralization, consensus algorithms, and cryptographic security are explored. There is also a detailed overview of smart contracts and their uses, along with discussing the potential advantages and disadvantages of blockchain technology  \cite{zheng2018blockchain}.

An innovative method of data sharing through blockchain technology is introduced. The study proposes a semi-decentralized approach that employs the use of InterPlanetary File System (IPFS) to facilitate secure and efficient data sharing. According to the proposed method, the data owner first uploads an encrypted file to IPFS, which is then divided into n secret sections known as hash codes. Next, the data owner sets the access permissions for the encrypted file by specifying seven access rights. This unique approach ensures that the data remains protected and can only be accessed by authorized parties. The authors of the study investigated an Ethereum public blockchain framework and found that it supports on-chain and off-chain transactions, cloud deployment, pseudonymity, access control, and consensus. However, the study did not address the limitations of the framework or the issue of interoperability with other blockchain systems \cite{athanere2022blockchain}.

 An evaluation of various blockchain platforms took place, leading to the conclusion that Hyperledger Fabric was the most suitable platform for their research objectives. The paper highlights the platform's strong privacy features, particularly its role-based access mechanism, which allows for the regulation of user access to data and transactions based on their designated role within the system. Additionally, Hyperledger Fabric utilizes a permissioned network, ensuring that only authorized nodes can participate in the network and access data. However, the study did not investigate the cloud deployment functionality of the framework or provide an extensive analysis of their system's security \cite{saleh2020blockchain} .

Introduction of a blockchain-powered platform named UniverCert for certificate verification is explored. This platform is built on the Ethereum blockchain technology in its consortium form, and its stakeholders include higher education institutions, governments, law enforcement agencies, and employers. The RestAPI channel is used for accessing the platform, providing users with convenient access to the platform's features. However, the authors did not address the issue of how their proposed system would handle situations where graduates may desire to keep their personal information private. Further exploration and discussion of these limitations could offer valuable insights for future research in this field \cite{shakan2021verification}.

 The paper follows an extensive discussion of the enhancement of digital document security through the implementation of timestamping and digital signatures. The digital signature comprises four fundamental components, namely, hash code, public key, private key, and timestamp. The university provides the student with both a printed copy of their educational certificate and a digitally signed document. Nonetheless, the authors did not investigate the on-chain, off-chain, cloud deployment, and consensus features of the framework or explore the issue of interoperability with other blockchain systems  \cite{ghazali2018graduation}.

Incorporating an additional accrediting body to the certification verification process is a crucial step towards ensuring the authenticity of universities authorized to issue and verify certificates. This feature adds an extra layer of validation, increasing the system's security and trustworthiness. To ensure the confidentiality and security of data, the AES encryption algorithm was employed. The system also allows for the submission and verification of multiple academic certificates simultaneously, streamlining the verification process and enhancing its efficiency, as highlighted by \cite{leka2021development}. The authors observed that the framework supports on-chain transactions, pseudonymity, and access control. However, they did not delve into the off-chain and consensus features of the framework or evaluate the performance of their system concerning latency and throughput.

 Utilization of the Go implementation of Ethereum, commonly referred to as GETH is used to establish a blockchain network that stores certificates. The study's reliability tests demonstrated that the system could handle roughly 200 transactions in eight seconds, indicating its effectiveness. The scalability tests indicated that to become a node or miner on the blockchain network, one would need a storage capacity of 22.6 GB to support ten million blocks. Although the system was comprehensive in its scalabilty, it failed to assess the access control mechanism of the system \cite{faaroek2022design}.

 MIT Media Lab uses Blockcerts to issue digital certificates to student groups, giving the certificate holders more control over their earned certificates. The certification process involves the issuer signing the digital certificate and storing its hash on the blockchain, with the recipient receiving the output. However, this process resulted in ownership issues and the need for a high level of trust. Furthermore, the MIT Media Lab has also released an app called Blockcerts Wallet, which stores information about diplomas and the key pairs of students \cite{vidal2019analysis} .

In an effort to enhance the efficiency of transaction throughput, Kafka was integrated into the message queuing process. The utilization of this technology resulted in faster transaction processing times, surpassing other blockchain-based solutions. The study used Hyperledger Fabric, a well-known blockchain framework used for creating enterprise-level applications. The researchers' findings are crucial as the faster processing times mean a boost in overall efficiency and decreased expenses for businesses. However, the study did not delve into the security of their system or investigate the cloud deployment feature of the framework \cite{liu2019blockchain}.

Exploration of several applications took place in this particular paper. Firstly, the researchers analyzed three of the most widely-used blockchain-based cryptocurrencies, which were Bitcoin, Litecoin, and Ethereum. Secondly, they explored the features and concerns surrounding the Bitcoin cryptocurrency. Finally, the team developed a graphical user interface for IPFS bandwidth analysis, which allows files to be stored on the network using Web3 JS and Smart Contracts. This research provided a better understanding of the blockchain technology and its different applications, revealing insights into the advantages and limitations of cryptocurrencies. The framework utilized pseudonymity to safeguard the privacy of supply chain participants and used on-chain transactions to document supply chain events and audit trails \cite{nouman2021secure}.

The articles reviewed cover a range of topics related to the use of blockchain technology for certificate verification, digital certificates, and smart contracts. The studies explore the benefits and limitations of blockchain technology, including its security, decentralization, and consensus algorithms. One of the key benefits of blockchain technology is its ability to provide increased security and trust in the verification process, particularly when multiple accrediting bodies are involved. Overall, the studies demonstrate the potential of blockchain technology to improve the efficiency and security of various processes, particularly those related to verification and certification. However, the authors also note the need for further research to explore the scalability, performance, and security of these blockchain-based solutions.

\section{Proposed System}\label{sec3}
This section outlines the proposed certificate verification system based on blockchain technology.To begin with, an applicant uploads their necessary credentials to the system. Once the credentials are uploaded, the admin gets a request from the applicant to verify the certificates. The admin reaches out to educational organizations to confirm the authenticity of these certificates. Once the credentials are verified, the admin notifies the users and uploads the certificates to IPFS. 

When students upload the certificates to an existing protocol node in IPFS, the data is chopped into smaller segments and undergoes a hashing algorithm. Once the hashing process is complete, it returns a hash key also known as the content identifier (CID). The CID serves as a fingerprint to uniquely identify the files. A new cryptographic hash (CID) is generated for each new upload of new data or previously uploaded data. This makes each upload to the network unique and resistant to security breaches or tampering. 

To enhance security, the hash key is encrypted before it is stored in the Blockchain nodes. After the data is transmitted to the Blockchain through IPFS, the issuer must approve the generation charges in Metamask. Typically, the hash cannot be modified once it is stored in the Blockchain. In the event of any data tampering, the system immediately alerts users.

The applicant can now send the hash number to organizations in order to apply for any job. The companies can use the system to type in the hash to look for the applicants and find the legitimacy of their certificates.  

\subsection{Proposed Framework}
\begin{figure}[htbp]
  \centering
  \includegraphics[width=0.6\textwidth]{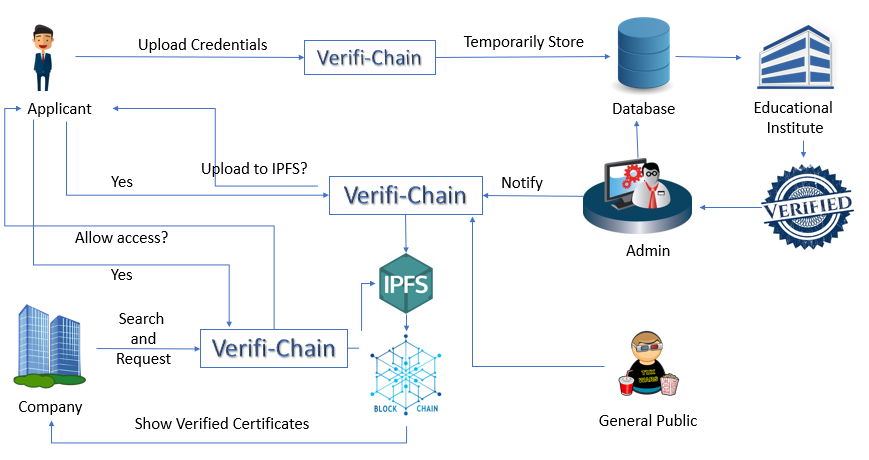}
  \caption{ Workflow of the system }
  \label{fig:example}
\end{figure}

Figure 1 explains the workflow of the system. At first, the user login to the system and upload their credentials. These certificates are temporarily stored in the database. The admin receives a notification from the user regarding the verification request. Once the admin verifies the certificates, these are uploaded to the IPFS. The company, on the other hand, can search the applicants and request access to solely view the certificates. Furthermore, the certificate is impervious to any further adjustments or revisions requested by an intruder.

\subsection{Stakeholders of The System}
In order to address the flaws in the current methodologies, this proposed system, both stores certificates and acts as a verifier. The entire procedure is thoroughly explained in the sections that follow.

By utilizing distributed technologies such as IPFS and Ethereum smart contracts, our proposed solution simplifies the process of verifying the authenticity, integrity, and validity of a document. The stakeholders involved in the system are as follows. 

\begin{enumerate}
\item \textbf{Applicant:} An applicant is able to upload their necessary credentials to the system. Once their credentials are verified, it is stored in the IPFS. Furthermore, applicants can accept access requests from companies to view their verified certificates. 
\item \textbf{Company:} A company is able to search for an applicant and request access to applicants in order to view their credentials. 
\item \textbf{Admin:} The admin receives verification requests from the applicants to confirm the originality of their certificates. The admin reaches out to educational organizations to confirm the validity of the certificates. 
\item \textbf{General User:} A general User is able to browse through the system's homepage and get acquainted with the functionality.
\end{enumerate}

\subsection{System Modules}
Blockchain can serve as the foundation for an entire project, with its immutable ledger system providing a secure method for storing data in Blockchain nodes. The blockchain is composed of a sequence of blocks, each containing records of multiple transactions. Once a block is added to the chain, the information it contains cannot be modified, ensuring the data's integrity is stored on the blockchain. This security is achieved through the use of cryptographic algorithms and a consensus mechanism that verifies transactions and maintains the ledger's integrity. The technology is designed to be highly secure and resistant to tampering and hacking, making it well-suited for storing valuable and sensitive information. Furthermore, it has numerous applications, including supply chain management, digital identity verification, voting systems, and more.

Ethereum is an open-source blockchain platform with smart contract functionality and is powered by its native cryptocurrency, ether (ETH). Smart contracts are programs that are executed on the blockchain when a specific action is performed by a user. They can be written in various programming languages, with Solidity being a popular choice. On the other hand, the Interplanetary File System (IPFS) is a decentralized file storage system that enables the creation of a peer-to-peer network among computers worldwide. Each file in the global namespace of IPFS is uniquely identified by content-addressing. IPFS relies on a network of nodes to store and distribute files, rather than depending on a central server. This decentralized architecture provides several benefits, including increased reliability, security, and speed. Whenever a file is uploaded or added to the IPFS network, it is assigned a unique identifier known as a hash. This hash serves as a digital fingerprint for the file and can be used to retrieve the file from any node in the network that has it stored.

In summary, Ganache, Metamask, Truffle, React, and Node JS are all tools that can be used together to create an effective Blockchain platform. Ganache acts as a local server for the Ethereum Blockchain, Metamask is a cryptocurrency wallet that tracks transactions on the Blockchain, and Truffle is a framework for compiling, linking, deploying, and managing smart contracts. React is a framework for developing user interfaces, while Node JS is used for serving frontend pages, and assets, and managing user authentication via JWT. Web3 is a dependency of Node JS that allows solidity code to be run on the front end.

\subsection{Technical Diagram}
\begin{figure}[h!]
  \centering
  \includegraphics[width=0.7\textwidth]{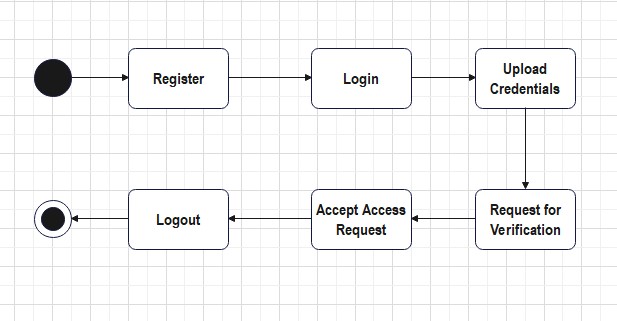}
  \caption{Applicant Activity}
\end{figure}

Figure 2 depicts the applicant's use case. In order to utilize the system, the applicant must first register and log in. The applicant can upload their credentials in the system and request the admin for verification. While the verification process is still ongoing, the certificates will be temporarily stored in a local database. When a company requests access to view the credentials, the applicants have the option to review the access request and accept it accordingly. 

\begin{figure}[ht]
  \centering
  \includegraphics[width=0.5\textwidth]{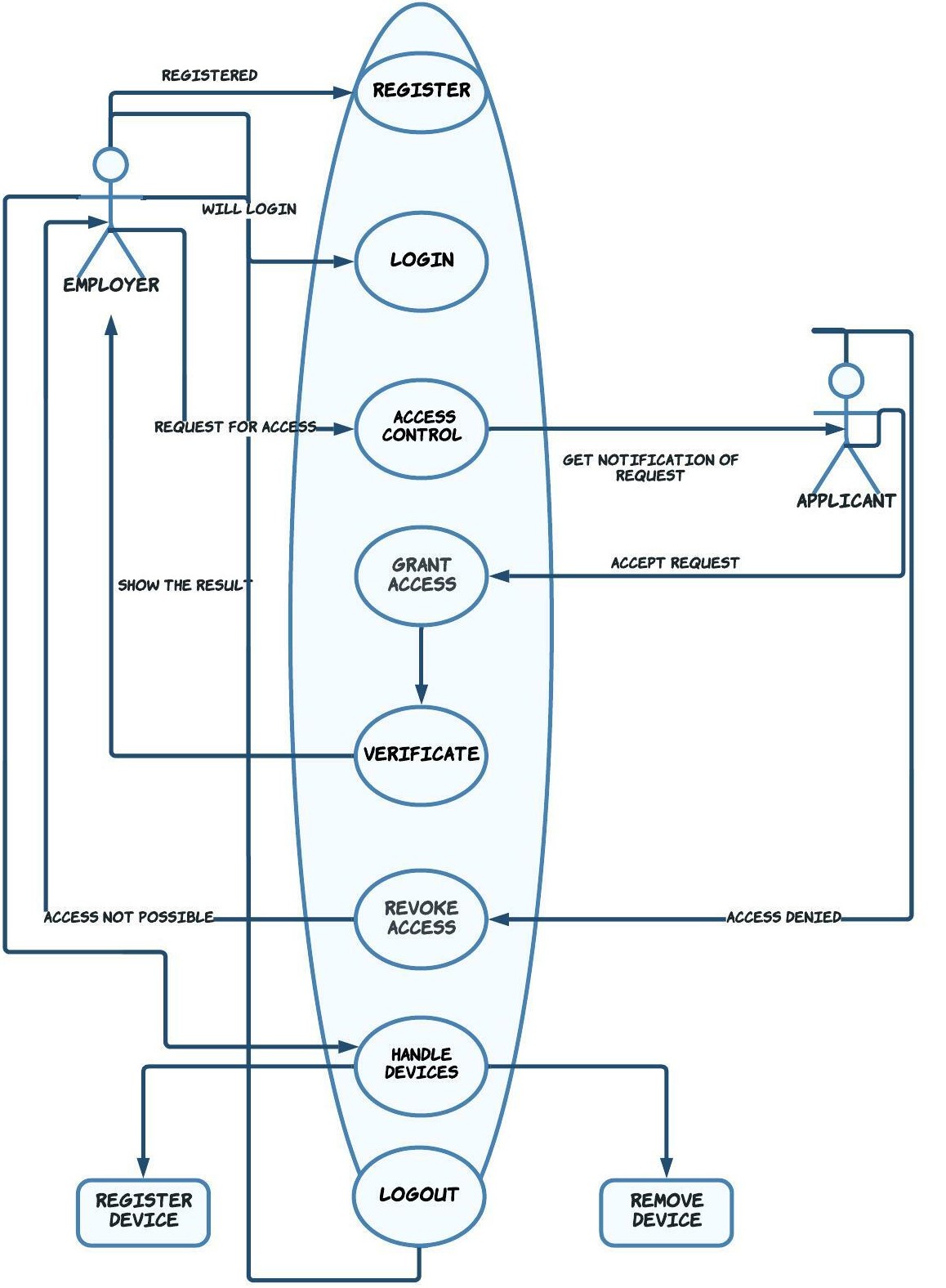}
  \caption{Use Case of Company}
\end{figure}

Figure 3 shows the use case for the company. When the firm receives the hash key from the applicant, it uses it to search the system for the specific applicant. Because the system has access tiers for further protection, the company must first request access before seeing the credentials. The company is able to examine the certificates after the applicant approves the request from their account.

\begin{figure}[ht]
  \centering
  \includegraphics[width=0.7\textwidth]{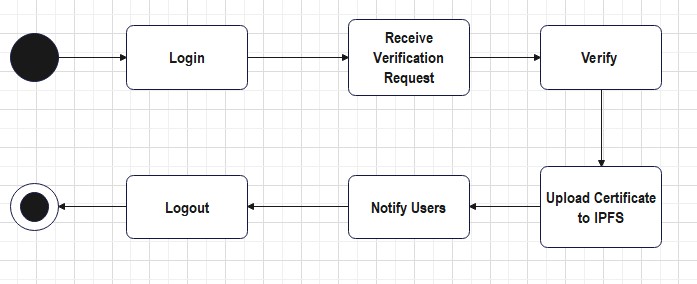}
  \caption{Activity of The Admin}
\end{figure}
Figure 4 illustrates the administrator's activity. The administrator's job is to receive verification requests from applicants and contact certificate providers such as educational institutions to check the certificates' legitimacy. They upload the confirmed results to the IPFS and notify the applicant once they get the results.

\begin{figure}[ht]
  \centering
  \includegraphics[width=0.7\textwidth]{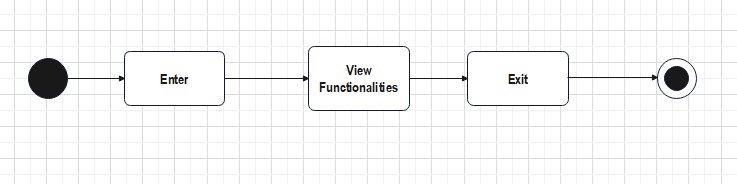}
  \caption{Activity Diagram for the General User}
\end{figure}
Figure 5 illustrates the activity for the general user. The general user can simply browse through the pages of the system and view the functionalities.

\section{Conclusion}
One of the primary advantages of Blockchain is the ability to create immutable ledgers. Due to the ever-growing rate of certificate falsification during the job application process, this proposed solution aims to store and validate academic certificates using blockchain and IPFS. The system not only makes it easier to verify certificates, but it also reduces the risk of losing tangible certificates by storing them digitally in IPFS. While the original document is kept in IPFS, the hash associated with the certificate is kept in the blockchain. Future work could explore the integration of other blockchain platforms and file storage systems further to enhance the functionality and robustness of the system. Overall, the proposed solution has the potential to revolutionize the way academic certificates are verified, making the process more transparent, efficient, and secure for all stakeholders involved.


\bibliography{sn-bibliography}


\end{document}